\documentstyle[11pt,newpasp,twoside,epsf]{article}

\begin{document}

\title{Multiwavelength analysis of Gl\,355 (LQ\,Hya)}

\author{S.~Covino$^1$, G.~Tagliaferri$^1$, E.~Bertone$^1$, G.~Cutispoto$^2$,
S.~Messina$^2$, M.R.~Panzera$^1$, L.~Pasquini$^3$, R.~Pallavicini$^4$,
S.~Randich$^5$, M. Rodon\'o$^2$, J.~Setiawan$^6$}
\affil{$^1$Brera Astronomical Obs., Via Bianchi 46, 23807 Merate, Italy}
\affil{$^2$Catania Astrophysical Obs., Via S. Sofia 78, 95123 Catania,
Italy}
\affil{$^3$ESO, Karl S. Strasse 2, 80457
Garching bei M\"unchen, Germany}
\affil{$^4$Palermo Astr. Obs., P.za del Parlamento 1,
90134 Palermo, Italy}
\affil{$^5$Arcetri Astrophysical Obs., Largo E. Fermi 5, 50125 Firenze,
Italy}
\affil{$^6$Kiepenheuer Inst. Sonnenphysik, Freiburg, Germany}

\begin{abstract}
We discuss ROSAT, ASCA, {\it Beppo}SAX and optical observations of the young
active star Gl\,355. During the ROSAT observation a strong flare was detected
with a peak flux more than an order of magnitude larger than the quiescent
level. Spectral analysis of the data allows us to study the temperature and
emission measure distribution, and the coronal metal abundance, for the
quiescent phase and, in the case of ROSAT, also during the evolution of the
flare. We have modeled the flare and derived a loop semi--length of the order of
$\sim 1.5$ stellar radii. ROSAT, ASCA and {\it Beppo}SAX data suggest that the
coronal abundance of Gl\,355 is subsolar, in the range $0.1 \div
0.3\,Z/Z_\odot$. A preliminary analysis of optical spectra allows us to compare
the photospheric and coronal metal abundances.
\end{abstract}

\section{Introduction}

We present here the analysis of X--ray and optical observations of Gl\,355
performed on a time--scale of 10 years with ROSAT, ASCA, {\it
Beppo}SAX, the 80cm Automated Photometric Telescope (APT)  in Catania
(Italy) and the ESO 1.5m spectrographic telescope. Gl\,355 is a relatively well
known nearby ($d \sim 18$\,pc) star of spectral type K2Ve. The high lithium
abundance and rotation rate suggest a young age, possibly that of a pre--main
sequence object (Vilhu et al. 1991) or more likely a ZAMS star. The high
count--rate of the ROSAT observations has allowed us to perform a time--resolved
spectral analysis of an intense flare to discuss the temporal evolution of
plasma parameters such as the temperature $T$, the emission measure $EM$, the
global coronal metallicity $Z$, and of the absorbing column density $N_{\rm H}$.
The quiescent emission was studied using ROSAT, ASCA and also {\it Beppo}SAX
observations and the long--term behavior was investigated comparing measurements
performed at different epochs. Optical monitoring of Gl\,355 was also performed
with the APT to compare the variability in the X--ray and optical band. High
resolution spectra obtained at ESO allows us to directly determine the
photospheric metal abundance, a parameter still lacking for this object, and to
compare in a more meaningful way its coronal and photospheric metallicities.
Part of the data here reported was already discussed in detail in Covino et al.
(2001).

\section{Observations and analysis}
\label{sec:observations}

The X--ray observations discussed in this paper were performed by the ROSAT,
ASCA, and {\it Beppo}SAX satellites in Nov 1992, May 1993, December
2000, respectively.

\begin{figure}[t]
\plotone{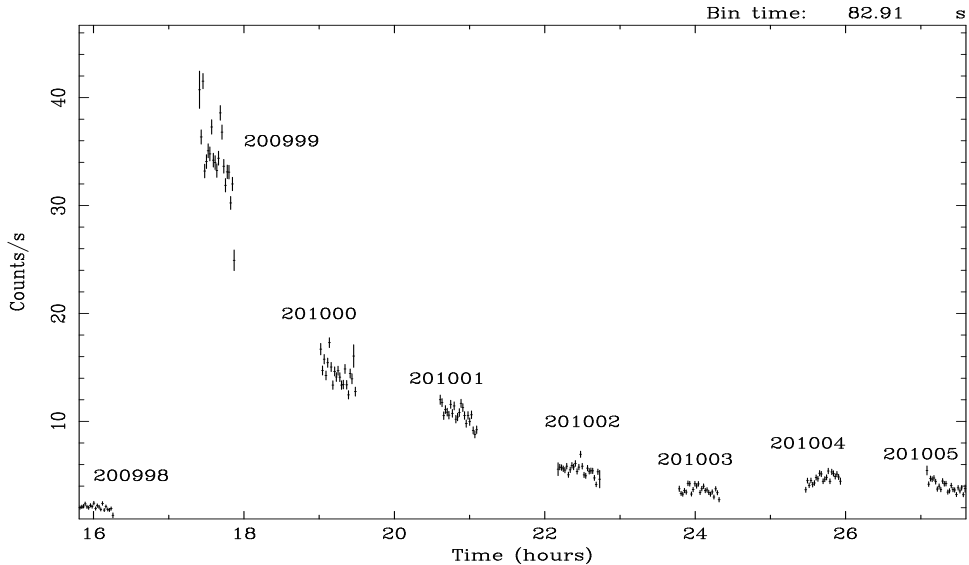}
\caption{Total light curve obtained by the ROSAT PSPC observations. The count
rate is not background subtracted. Observations were performed starting on 1992,
November 5.}
\label{fig:totlc}
\end{figure}

The ROSAT light curve shows an evident flare which occurred on 1992, Nov 5, at
$\sim 18$ UT. The count rate increases by more than an order of magnitude.
Fig.\,\ref{fig:totlc} shows the complete light curve with superimposed the
identification codes of the specific pointings. The flare maximum can
be located close to the observation 200999 or just before it.

The ASCA count rate is essentially constant for the first half of the
observation, $\sim 0.4$\,cts for the SIS. The second half, on the contrary,
shows an increase of $\sim 50$\% in particular in the softest band. However, the
hardness ratio does not show significant variations due to the relatively poor
statistics involved.

\begin{figure}[t]
\plotone{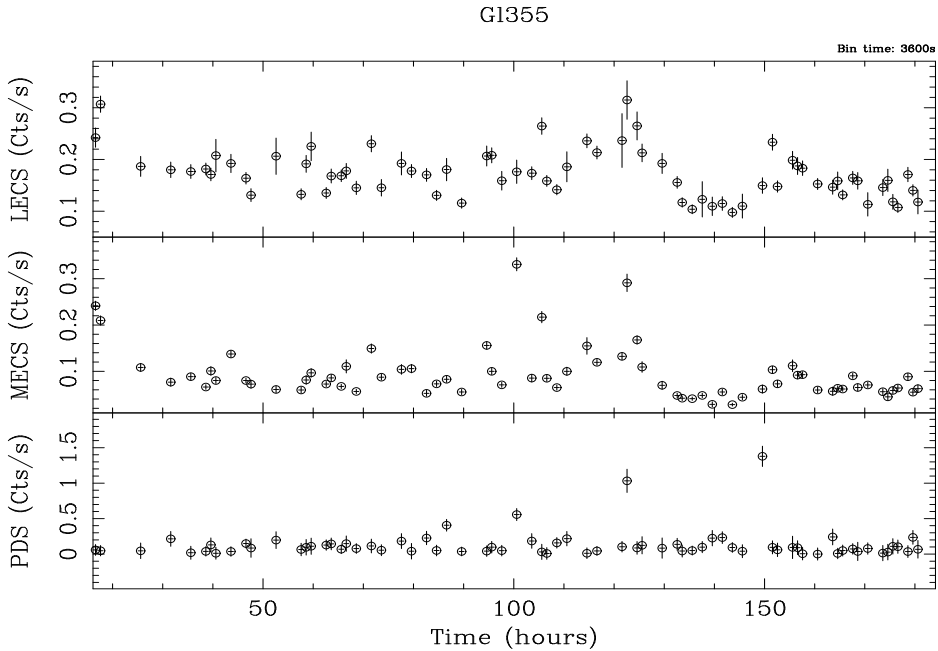}
\caption{Total light curve obtained with the {\it Beppo}SAX LECS, MECS and PDS
detectors. Observations cover the time--interval from December 5 to 12, 2000.}
\label{fig:saxlc}
\end{figure}

The  {\it Beppo}SAX observation lasted for 280\,ks and allows us to study in
detail the short term variability of this object. No intense flares were
detected even if a continuos low--level activity is clearly present
(Fig.\,\ref{fig:saxlc}).

\begin{figure}[t]
\plotone{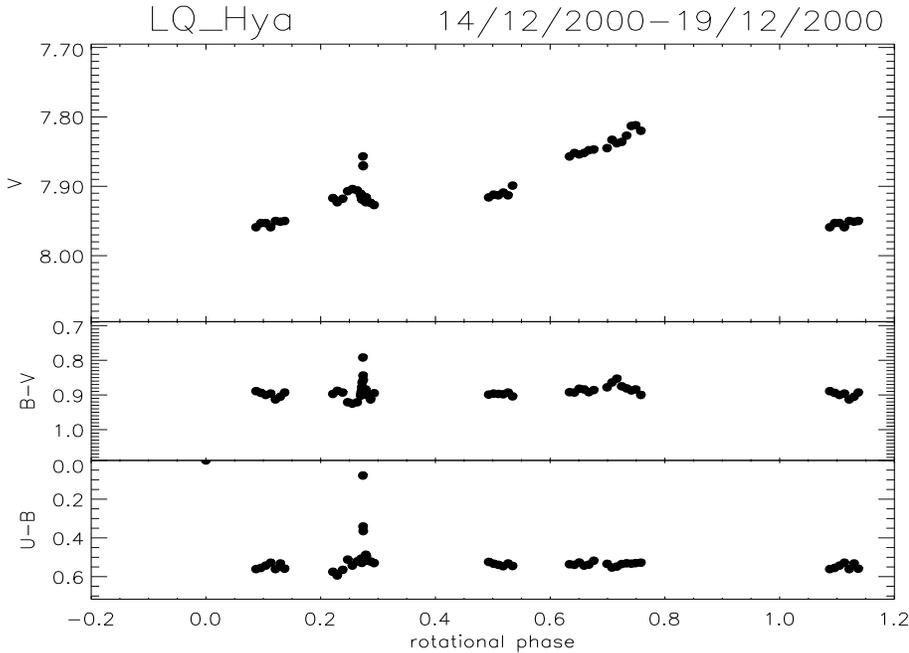}
\caption{A flare observed in the optical. The flare evolution is superposed to
a general trend likely due to spots on the star surface.}
\label{fig:flopt}
\end{figure}

Optical monitoring of Gl\,355 was performed from December 2000 to March 2001.
The observations were performed with the APT  of Catania Observatory (Mt.\,Etna,
Italy). It is a 0.8-m reflector equipped with an uncooled Hamamatzu R1414 SbCs
phototube and standard UBV filters. Apart from a long term variability a flare
in the optical was also observed (Fig.\,\ref{fig:flopt}).

A high-resolution spectrum was obtained on Nov. 3rd, 2000 with the FEROS
spectrograph (Kaufer et al. 1997) at the ESO 1.5m telescope. The spectrum covers
the entire optical band from 370 to 920\,nm with a resolving power of R =
48,000. A detailed abundance analysis is in
progress; preliminary results suggest that the global photospheric
metallicity of Gl\,355 is slightly over-solar.

\section{Results}
\label{sec:results}

\subsection{Quiescent Emission}
\label{sec:quie}

The ROSAT, ASCA and {\it Beppo}SAX observations and the
RASS data allow us to study the quiescent emission of Gl\,355 on long and 
short time--scales. Long-- and short--term variability is clearly present
amounting up to a factor of $2 \div 3$ in flux.

The ROSAT PSPC observations outside of the flare were fitted with a 1T model
with free global metallicity. The absorbing column across the line of sight is
not well constrained (errors of the order of 50\% of the best fit value, or
larger). The metal abundance is highly subsolar ($Z/Z_\odot = 0.03 \div 0.14$ at
the 90\% confidence level) and is confirmed by the analysis of the ASCA spectra.
The best--fit temperature is $\sim 0.7$\,keV. The flux in the 0.1--2.4\,keV
energy band ranges from $\sim 1.4$ to $\sim 3.5 \times
10^{-11}$\,erg\,cm$^{-2}$s$^{-1}$, directly linked to the EM variations from
$\sim 9$ to $\sim 21 \times 10^{52}$\,cm$^{-3}$. The spectral parameters derived
for the pre-- and post--flare emissions are comparable and the moderate (a
factor of $2 \div 3$) flux variability seems to be due mainly to EM variations
which in turn might be due to changes in either the volume or the density of the
emitting regions.

\begin{figure}[t]
\plotone{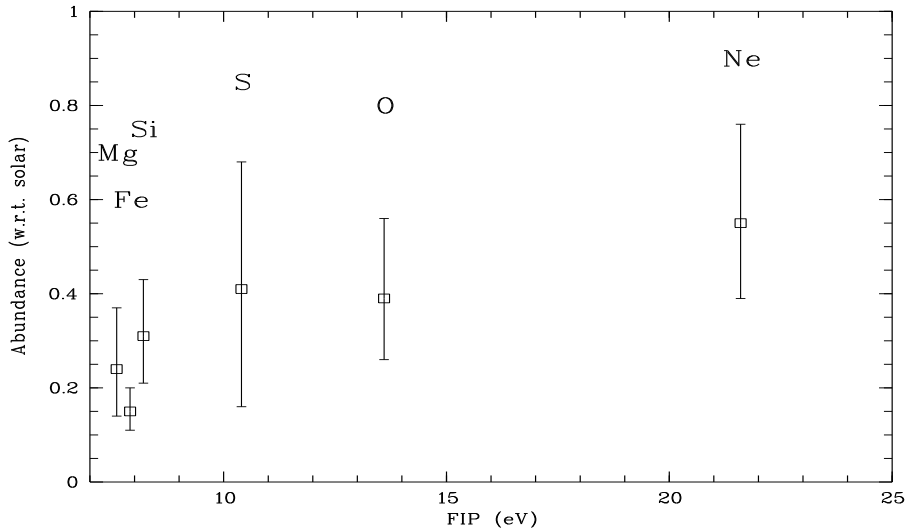}
\caption{Comparison of metal abundance derived by ASCA data and FIP for single
elements. The error bars are large but a possible trend for an inverse FIP
effect seems to be present.} \label{fig:trend}
\end{figure}

Considering the ASCA observation on May 1993 1T spectral fits gave satisfactory
results only for the low resolution GIS detectors. The best fit temperature
appears slightly harder (but comparable within the errors) than the one derived
from the ROSAT analysis, while the metal abundance is essentially the same.
Acceptable fits to the SIS spectra were obtained only with the addition of a
second thermal component, with the global metallicity free to vary. An analysis
of the single element abundance shows (Fig.\,\ref{fig:trend}) a possible inverse
First Ionization Potential (FIP)--effect,  but the error bars are large (see Covino et al. 2000 for a brief
discussion).

{\it Beppo}SAX data analysis is still preliminary, but reveals a complex 3T
structure thanks to the wider energy range. The temperatures are rather well
constrained ($T_1 = 0.25 \div 0.45$, $T_2 = 0.9 \div 1.1$, $T_3 = 2.3 \div
3.5$\,keV) and the metal abundance is again strongly subsolar, around
$0.2 \div 0.3\,Z/Z_\odot$.

Therefore,the metal abundance measured by ASCA and {\it Beppo}SAX
 is comparable to that obtained from the analysis of the ROSAT data
and is well below solar. Since ASCA and {\it Beppo}SAX are much more effective
than ROSAT in measuring metal abundances, this result clearly shows that a low
metal abundance is indeed needed to model the corona of Gl\,355 in contrast with its young
age and about solar photospheric metallicity (see Sect(s).\,\ref{sec:observations} and \ref{sec:concl}).

\subsection{Flare analysis and modeling}
\label{sec:flare}

Considering the ROSAT observations 200999, 201000, 201001 and 201002, i.e. the
observations performed during the flare, we performed 2T fits with the global
metallicity left free to vary. The fits gave acceptable results, but the hotter
component is badly constrained. The intense dynamic evolution of the flare
prevents a detector with a limited energy band as the ROSAT PSPC to constrain
the hot component.

\begin{figure}[t]
\plotone{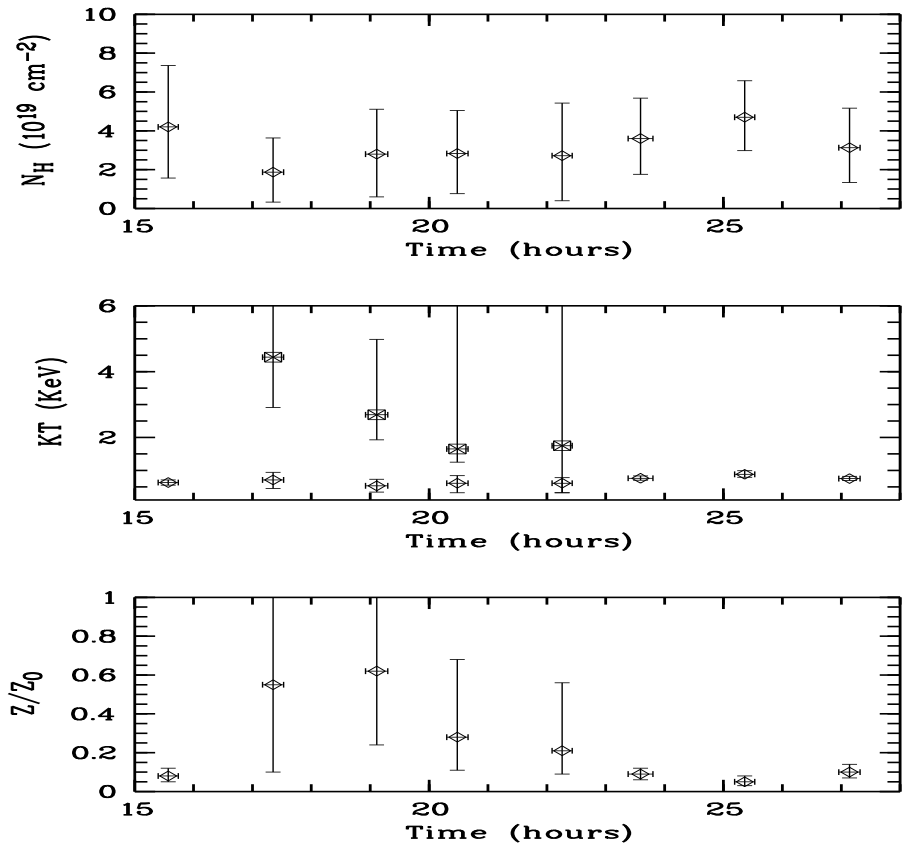}
\caption{1T and 2T best fit parameters for the ROSAT PSPC observations. The
first observation and the last three were fitted by 1T models, while the
observations performed during the flare were fitted with 2T models
(see Fig.\,\ref{fig:totlc}).}
\label{fig:T}
\end{figure}

As shown in Fig.\,\ref{fig:T}, the temperature of the flare cooler component is
essentially constant around 0.6--0.7\,keV, in full agreement with the
temperature derived for the quiescent emission. The absorbing column density is
also essentially constant during the flare, i.e. with no increase at the flare
onset or at the peak. The metal abundance close to the flare peak is not well
constrained but a hint (admittedly very weak) for an increase during the flare
evolution seems to be present. In any case a hot component is clearly needed,
although being not well constrained due to the limited ROSAT energy range. The
flare event is totally due to hot plasma superimposed to the quiescent corona.

The flare observed by ROSAT has been analyzed considering the so called
hydrodynamic decay--sustained heating scenario (Reale et al. 1997),
which assumes that the flaring plasma is confined in a closed loop structure
whose geometry does not change during the event. Detailed hydrodynamical
simulations (Peres et al. 1982, Betta et al. 1997) show that flares decay
approximately along a straight line in the $\log \sqrt{EM} - \log T$ diagram,
and that the value of the slope $\zeta$ of the decay path is related to the
ratio between the observed decay time $\tau_{\rm lc}$ of the light curve and the
thermodynamic cooling time of the loop $\tau_{\rm th}$ in the absence of heating
during the flare decay. This allows deriving the length of the flaring loop as a
function of observable quantities. An application of this technique to stellar
flares observed with the ROSAT PSPC, and the appropriate calibrations, are given
by Reale \& Micela (1998) and Favata et al. (2000). See also Pallavicini et al.
(2000) and Covino et al. (2001) for a discussion of the method.

In the present case, the slope $\zeta$ in the $\log \sqrt{EM} - \log T$ is $0.86
\pm 0.27$. Very low values of $\zeta$ mean that the flare decay is entirely
driven by the sustained heating, so that the thermodynamic cooling time,
$\tau_{\rm th}$, cannot be determined in a reliable way, and $L$ does not depend
any more on $\tau_{\rm lc}$. On the other extreme, no sustained heating occurs
and $\tau_{\rm lc} \sim \tau_{\rm th}$. Our result shows that indeed a large
amount of sustained heating is actually present in line with the results
obtained for other flares (Reale \& Micela 1998; Ortolani et al. 1998, Favata \&
Schmitt 1999; Favata et al. 2000a; Favata et al. 2000b; Favata et al. 2000c;
Maggio et al. 2000, Franciosini et al. 2001).

Given the decay e--folding time, $\tau_{\rm lc} = 10.1 \pm 0.5$\,ks, the loop
semi--length turns out to be $L = 83 (\pm 22) \times 10^{9}$\,cm or, in units of
the stellar radius, $L \sim 1.5\,R_*$.

\section{Summary and conclusions}
\label{sec:concl}

The spectral analysis of the ROSAT, ASCA and {\it Beppo}SAX data shows that this
young star has coronal metal abundances highly sub--solar. Preliminary abundance analysis of the optical
spectrum indicates a metallicity close to solar.
Thus, as in the case of other young objects, e.g. AB\,Dor (Mewe et al.
1996) and HD\,35850 (Tagliaferri et al. 1997), we have a star that shows coronal
abundances much lower than the photospheric ones. This is also true for other
more evolved stars, as e.g. II\,Peg (Mewe et al. 1997; Covino et al. 2000).
These ROSAT, ASCA and {\it Beppo}SAX findings are now confirmed with the
gratings observations of Chandra for HR\,1099 (Drake et al. 2001) and of
XMM-Newton for AB\,Dor, Castor and HR\,1099 (Brinkmann et al. 2001, G\"udel et
al. 2001a, 2001b).

Besides the strong flare detected by ROSAT, variability of a factor of 2-3 has
been detected in both ROSAT, ASCA and {\it Beppo}SAX light curves of Gl\,355.
This variability is mainly due to EM changes, while the temperature of the
quiescent corona remains approximately constant. For the ROSAT data, outside the
flare, the coronal plasma is well represented by a single temperature model with
very low metal abundances. For the ASCA data, due to the
harder energy band, a second component is required while the wider {\it
Beppo}SAX energy range also requires a third component.

The coronal spectrum during the flare can be represented with a 2T model, with
the cooler component compatible with the quiescent one. From these fits we have
a weak indication of an increase of the metal abundance during the flare
although the large error bars do not allow for a strong claim.

We modeled the flare using the hydrodynamic decay--sustained heating scenario
(Reale et al. 1997) and assumed that in the 2T best fit model the
hotter temperature represents the flare plasma, while the cooler temperature
represents the quiescent coronal plasma. We then derived the flare loop
semi--length that turns out to be quite large, $\sim 1.5\,R_*$. Note that our
flare temperature (the hotter component) in not well constrained at higher
values, due to the ROSAT energy band. In any case, since the downward error bars
for the temperature are well constrained by the fit, the loop semi--length that
we derived should be regarded as a lower limit.

\acknowledgements
This work has been partially supported by the Italian Space Agency (ASI)

\end{document}